\documentstyle[aps]{revtex}
\begin{document}
\title{\bf $^1$S$_0$ pairing correlations in relativistic nuclear matter
and the two-nucleon virtual state}

\author{B. V. Carlson and T. Frederico}
\address{
Departamento de F\'{\i}sica, Instituto Tecnol\'ogico da Aeron\'autica -- CTA,
12.228-900 S\~ao Jos\'e dos Campos, S\~ao Paulo, Brazil}

\author{F.B. Guimar\~aes}
\address{
EAN -- Instituto de Estudos Avan\c cados -- CTA,
12.228-840 S\~ao Jos\'e dos Campos, S\~ao Paulo, Brazil}

\maketitle

\begin{abstract}
We use the Gorkov formulation of the Dirac-Hartree-Fock-Bogoliubov
approximation to nuclear pairing to study the $^1$S$_0$ nucleon-nucleon
correlations in nuclear matter. We find the short-range correlations of
the $^1$S$_0$ pairing fields to be almost identical to those of the
two-nucleon virtual state. We obtain mutually consistent results for
the pairing fields, using several different sets of effective interaction
parameters, when we demand that each of these sets places the virtual-state
pole at its physical location.
\end{abstract}

\pacs{PACS number(s): 21.65.+f, 74.20.Fg, 21.60.Jz }

The non-relativistic BCS and Hartree-Fock-Bogoliubov (HFB) approximations
have been used with success to study pairing in nuclear physics, both years
ago\cite{bmp,em1,em2,ems,saufess,fessau,kagun} and still today
\cite{kringsh,baldo1,baldo2,baldo3,clark,baldo3p,baldo4,khodel,nordic}. 
Pairing approximations provide a simple means of extending the 
independent-particle approximation to one which includes the effects of the
binding energy and short-range correlations associated with bound pairs of 
nucleons in the nuclear medium. The long-range correlations associated with 
collective phenomena in nuclei are similarly described using the 
non-relativistic random-phase approximation (RPA).

On the other hand, it is well known that non-relativistic independent-particle
approaches using realistic two-body interactions have difficulties in
accounting for basic phenomena, such as the spin-orbit part of the
nucleon-nucleus interaction and the saturation properties of nuclear matter.
These properties can be fairly easily described in a relativistic formulation,
in which effective mesons are exchanged between Dirac nucleons. The success
of the relativistic mean field approach, initially developed by Walecka and
collaborators and later by many others
\cite{sw,hs-hf,bmgm,pw,hs-bruck,haarm,mach1,mach2}, invited its extension
to approximations that could take into account the residual correlations
between nucleons. 

Consistent relativistic formulations of both the RPA and the
HFB\cite{bailuv,cf,kring,gcf} have thus been developed. Of the latter, 
the authors of Ref. \cite{bailuv} derived self-consistent equations for the
components of the relativistic pairing field but performed no calculations.
The results obtained in Refs. \cite{cf} and \cite{kring}, using a zero-range
model and a non-relativistic reduction of the pairing equations, respectively,
are generally much larger than those obtained in non-relativistic calculations,
although both of the relativistic calculations used interaction parameters
that were fit to yield the saturation point of nuclear matter. Such results
would indicate that 
other aspects of the interaction are involved in pairing than those 
necessary to describe the bulk characteristics of nuclei and of nuclear matter.
The crudeness of the calculations, however, do not permit a firm conclusion
to be reached. The results obtained in the fully relativistic finite-range
calculations of Ref. \cite{gcf} are very similar and do permit such a
conclusion. Here we will show that, in addition to describing the bulk
properties of nuclear matter, a nucleon-nucleon interaction must provide
a good description of the 
two-nucleon $^1$S$_0$ virtual state, if it is to furnish a good description of
$^1$S$_0$ pairing in nuclear matter. 

The HFB approximation introduces short-range two-nucleon correlations 
through the pairing fields and their momentum dependence. These correlations
have been investigated in recent years in non-relativistic models of nuclear
pairing.\cite{baldo3,baldo4,khodel,nordic} In particular, the close
association 
between the short-range correlations of the two-nucleon $^1$S$_0$ virtual
state and those of the $^1$S$_0$ pairing fields was investigated in Ref. 
\cite{khodel}. We investigate the same association in the relativistic
model and reach a similar but stronger conclusion: the short-range
correlations of the pairing field are almost identical to those of the
two-nucleon virtual state. For the
relativistic interactions we have used, the short-range repulsion and
medium-range attraction resulting from the exchange of effective mesons 
provide a description of the short-range correlations in very good agreement
with those of the two-nucleon virtual state, deduced from realistic 
nucleon-nucleon potential models. By requiring that the interactions also
reproduce the position of the virtual state pole or, equivalently, the
two-nucleon singlet scattering length, we will show that the effective
relativistic interactions can provide a consistent description of pairing,
both among themselves and with the non-relativistic calculations.

\section*{Pairing and two-nucleon correlations}

In a two-particle system, bound-state correlations can be 
roughly classified as
asymptotic ones and short-range ones. The asymptotic ones are determined
principally by the binding energy, while the short-range ones depend on
the high-momentum components of the wave function. This can be seen by
examining the manner in which a bound pair appears in the two-body T-matrix,
$T(E)$. The T-matrix satisfies the integral equation
\begin{eqnarray}
T(E)=V+V\,G_0(E)\,T(E)\,,\label{tmat}
\end{eqnarray}
where $V$ is the two-body interaction and $G_0(E)$ the free two-body
propagator.
A bound state appears in the T-matrix as a pole at a negative value of the
energy, $E=-\epsilon_b$, where $\epsilon_b$ is the binding energy of the 
pair. We then have
\begin{eqnarray}
T(E) = d \frac{1}{E+\epsilon_b} d^{\dagger} +T_c(E)\,,\label{tbnd}
\end{eqnarray}
with $T_c(E)$ the continuum (positive energy) part of the T-matrix and 
$d$ the bound-state vertex function.
Substitution of the latter expression into the integral
equation immediately yields an equation for the vertex function,
\begin{eqnarray}
d=V\,G_0(-\epsilon_b)\,d\,.\label{bnd}
\end{eqnarray}
The vertex function describes the momentum dependence
(spatial dependence) of the bound state and is closely related to its 
wavefunction. By rewriting the equation for the vertex function as
\begin{eqnarray}
d = ( -\epsilon_b-H_0 )\,G_0(-\epsilon_b)\,d = V\,G_0(-\epsilon_b)\,d\,,
\end{eqnarray}
we can identify the bound-state wave function as 
$\psi=\,G_0(-\epsilon_b)\,d$.
In this last expression, we can see the rough division of the correlations
into asymptotic ones determined by the singularity of the Green's function,
$G_0(-\epsilon_b)$,
and short-range ones contained in the high momentum components of the vertex
function, $d$.

The two-nucleon system in the vacuum has a bound state in the isospin one,
$^3$S$_1$-$^3$D$_1$ channel --- the deuteron. In the isospin zero, $^1$S$_0$ 
channel, where we will study pairing, the two-nucleon system in the vacuum
has no bound state. It does have a virtual state, however, with 
$k_v\,\approx -0.05 i$ fm$^{-1}$ and $\epsilon_v= k_v^2/M \approx 140$ keV,
corresponding to a two-nucleon singlet scattering length
$a_0=1/|k_v|\,\approx 23$ fm. 
An expansion similar to that of Eq. (\ref{tbnd}) can still be
performed to extract the contribution to the T-matrix of the virtual state.
Because the energy of this state is extremely small, its contribution to the 
T-matrix dominates the low-energy scattering and essentially determines the 
short-range correlations in the $^1$S$_0$ channel. 
The short-range correlations determined by the momentum dependence of the
vertex function of the virtual state, $d_v(q)$, are thus to good
approximation given by the momentum dependence of the half-on-shell $^1$S$_0$
T-matrix at zero energy, $t_{10}(q,0;0)$. We have from
the expression analogous to Eq. (\ref{tbnd}),
\begin{eqnarray}
t_{10}(q,0;0)=<q|T_{10}(E=0)|k=0>\approx <q|d_v> 
 \frac{1}{\epsilon_v} <d_v|0>\,.
\end{eqnarray} 
Note that the asymptotic properties of the virtual state also play a role
here, by determining the magnitude of the contribution of the virtual
state to the T-matrix through the factor $1/\epsilon_v$. The
short-range correlations contained in the $q$-dependence can be
extracted unambiguously, however,
by normalizing both sides of the equation to their values at $q=0$.

To look for general two-particle correlations in nuclear matter, one could
study the Brueckner ${\cal G}$-matrix. In this generalization of the
two-particle T-matrix of Eq. (\ref{tmat}), the propagator 
$G_0(E)$ is now a many-body operator that, beside describing the 
two-body propagation, must take into account the effects of Pauli blocking
and of interaction with the nuclear medium. To look for bound-state
correlations, one would examine the poles and respective vertex functions of
the ${\cal G}$-matrix as advocated years ago by Emery\cite{em1,em2}.
Such an approach has been used by various
authors\cite{dick,vgdick,almro}. We will use instead the HFB
approximation. It too has been used to study bound-state
correlations in nuclear matter \cite{baldo3,khodel,nordic} and to even
estimate deuteron production in heavy-ion collisions \cite{baldo4}.

We can lend force to our use of the HFB approximation to study bound-state
correlations by analyzing
qualitatively the manner in which the one-body HFB approach takes
into account the two-body pairing bound state. We begin by observing that
a bound-state vertex function $d$ of the T- or ${\cal G}$-matrix 
can be considered as an operator that converts a two-body bound state
into two single particles. We could instead consider this a
one-body operator by (1) neglecting the particle number of the bound state
and (2) considering one of the outgoing particles as an entering hole. 
The effect of the vertex function would then be the conversion of a hole
to a particle. This is just what the pairing field, $\Delta$,
of the HFB approximation does. 
Analogously, we can associate the adjoint vertex function, $d^{\dagger}$, 
which converts two single particles to the two-body bound state, with the
conjugate pairing field, $\bar\Delta$, which converts a particle to a hole.

To obtain a complete one-body description, we must take into account the
propagation of particles and holes as well as the conversion of one to
the other. A HFB formalism that succeeds in unifying
these ingredients simply and clearly is the Gorkov one\cite{gor}.
Although infrequently used in non-relativistic studies of pairing,
the formalism has served as the basis for various relativistic studies
\cite{bailuv,kring,gcf}. This is due, at least in part, to its
natural expression in terms of the propagator
language common to field-theoretical approaches. 

The Gorkov formulation of pairing extends the usual particle
propagator $G(x-x')$ to one of the form
\begin{eqnarray}
\left( \begin{array}{cc}  
G(x-x') & F(x-x')\\ 
\widetilde F(x-x') & \widetilde G(x-x')
\end{array} \right) \,, \nonumber
\end{eqnarray}
in which $\widetilde G(x-x')$ describes the propagation of holes in the medium
and the anomalous propagator $F(x-x')$ and its conjugate $\widetilde F(x-x')$
describe the conversion of holes to particles and particles to holes,
respectively. Inverting the reasoning that lead us from the vertex function
$d$ to the pairing field $\Delta$, we can interpret the anomalous
propagator and its conjugate as terms describing the overlap between
the two-particle bound state and the two single particles. We thus expect
the anomalous propagators to contain information about the relative motion
of the two particles in the bound pair.

\section*{The Gorkov Formalism}

We sketch here the development of a Dirac version of Gorkov's 
self-consistent particle-hole propagator\cite{gcf}. To do this, we begin with
the following ansatz to the effective single-particle Lagrangian,
\begin{eqnarray}
\int dt  L_{eff} = \int d^4 x d^4 x' \left\{ \overline\psi (x)
(i \gamma_{\nu} \partial^{\nu} - M) \psi (x) \delta (x - x') +
\mu\overline\psi (x) \gamma_0 \psi (x) \delta (x - x')\right . \nonumber \\
\left . - \overline\psi (x) \Sigma (x - x') \psi (x') + \frac{1}{2}
\overline\psi (x) \Delta (x - x') \psi_T (x') + \frac{1}{2} \overline\psi_T
(x) \bar\Delta (x - x') \psi (x') \right \}\,,
\end{eqnarray}
where $\Sigma$ is the usual self-energy, $\Delta$  and $\bar\Delta$ are the
pairing fields and $\mu$ is a chemical potential, which will be used
to constrain the average baryon density.

The hole wave function, $\psi_T$, is defined as
\begin{eqnarray}
\psi_T=B\overline\psi^T\qquad\qquad\qquad\qquad
\overline\psi_T = \psi^T B^{\dagger}\,,
\end{eqnarray}
where $\psi^T$ denotes the transpose of the wave function $\psi$,
and the matrix $B=\tau_2\otimes\gamma_5C$, in which
the Pauli matrix $\tau_2$ acts in the isospin space and $C$ is the
charge conjugation matrix. We note that, up to a factor
of $\gamma_0$ and a phase, the isospin doublet $\psi_T$ is the
time-reverse of $\psi$.
  
The requirement that the effective Lagrangian be Hermitian yields the
following conditions on the self-energy and pairing fields,
\begin{eqnarray}
\Sigma(x)=\gamma_0\Sigma^{\dagger}(-x)\gamma_0\qquad\qquad\mbox{and}
\qquad\qquad\Delta(x)=\gamma_0\bar\Delta^{\dagger}(-x)\gamma_0\,.\label{herm}
\end{eqnarray}

The requirement of invariance under transposition of the pairing terms
yields the additional conditions
\begin{eqnarray}
\Delta(x)=-B^T\Delta^T(-x)B^{\dagger}\qquad\qquad\mbox{and}\qquad\qquad
\bar\Delta(x) = -B\bar\Delta^T(-x) B^*\,.\label{tran}
\end{eqnarray}
These constraints are important in limiting the possible structure of
the self-energy and pairing fields.
  
Making use of the relation between $\psi_T$ and $\overline\psi$, we can
manipulate the effective one-particle Lagrangian into a matrix form,
{\footnotesize
\begin{eqnarray}
\int dt L_{eff} = \frac{1}{2} \int d^4 x d^4 x'\;\;\left(\overline\psi (x) ,
\overline\psi_T (x)\right)\times\qquad\qquad\qquad\qquad\qquad\qquad\qquad
\qquad\qquad\qquad\qquad\qquad
\end{eqnarray}
\begin{eqnarray}
\left( \begin{array}{cc}
(i\gamma_{\nu}\partial^{\nu}-M+\mu\gamma_0)\delta(x-x')-
\Sigma(x-x') & \Delta(x-x') \\ 
\bar\Delta(x-x') & (i\gamma_{\nu}\partial^{\nu}+M-\mu\gamma_0)
\delta(x-x')+\Sigma_T(x-x'))
\end{array} \right)
\left( \begin{array}{c} \psi(x') \\ \psi_T(x')
\end{array} \right) \,, \nonumber
\end{eqnarray}}
where
\begin{equation}
\Sigma_T(x)=B\Sigma^T(-x) B^{\dagger}\,.
\end{equation}
The extended vector wave function is that of the quasi-particle. Its
equation of motion is evident and could be written down immediately.

Due to the translational invariance of nuclear matter, the equation in 
momentum-space is diagonal in the wave number. For the momentum-space
representation of the corresponding generalized Feynman propagator,
the equation of motion takes the form
\begin{eqnarray}
\left( \begin{array}{cc}
\gamma_{\nu} k^{\nu} - M - \Sigma (k) + \mu\gamma_0 & \Delta (k) \\
\bar\Delta (k) & \gamma_{\nu} k^{\nu} + M + \Sigma_T (k) - \mu\gamma_0
\end{array} \right)
 S_F (k) = 1\,. \label{eqmot}
\end{eqnarray}
We note that the momentum-dependent unitary transformation that diagonalizes
the matrix operator is the Dirac version of the Bogoliubov-Valantin
transformation.
  
To make contact between the effective quasi-particle Lagrangian and
an interacting one, we assume that the meson+interaction terms in the
latter have been reduced to four-fermion terms of the following form,
\begin{eqnarray}
\int dt  L_I = \frac{1}{2} \int d^4 x d^4 x' \overline\psi(x) 
\Gamma_{\alpha}(x)
\psi(x)D^{\alpha\beta} (x-x')\overline\psi(x')\Gamma_{\beta}
(x')\psi(x')\,,
\end{eqnarray}
where $\Gamma_{\alpha}(x)$ and $ \Gamma_{\beta}(x')$ are vertex functions,
$D^{\alpha \beta}(x-x')$ is the meson propagator, and $\alpha$ and $\beta$ 
represent any indices necessary for the correct description of the meson 
propagation and coupling. 

We can obtain
the mean field contribution of this interaction term by replacing each
of the possible pairs of fermion fields by its ground-state expectation value,
\begin{eqnarray}
\int dt (L_I)_{eff}
= \frac{1}{2}
\int d^4 x d^4 x'D^{\alpha\beta} (x - x')
\bigl\{ & &  2 \overline\psi (x) \Gamma_{\alpha} (x) \psi (x) 
<\overline\psi (x')
\Gamma_{\beta} (x') \psi (x') > \nonumber \\
& + & 2 \overline\psi (x) \Gamma_{\alpha} (x) < \psi (x) \overline\psi (x') >
\Gamma_{\beta} (x') \psi (x') \nonumber\\
& - & \overline\psi (x) \Gamma_{\alpha} (x) < \psi (x) \psi^T (x') >
\Gamma_{\beta}^T (x') \overline\psi^T (x') \nonumber \\
& - & \psi^T (x) \Gamma_{\alpha}^T (x) <\overline\psi^T (x) 
\overline\psi (x') >
\Gamma_{\beta} (x') \psi (x') \bigr \} \nonumber \,,\\
\end{eqnarray}
where, by $<...>$, we mean the time-ordered ground-state expectation value,
$<0\mid T(...)\mid 0>$.
  
We note that the first term in this expression is a Hartree one and the
second a Fock exchange one, while the last two, after using
the definition of $\psi_T$ to replace the transposed $\psi$'s, can be
recognized as pairing terms. Comparing these mean field contributions to
those of the effective quasi-particle Lagrangian, we can express
the self-energy and pairing fields in terms of the two-fermion ground-state
expectation values as
\begin{eqnarray}
\Sigma (x - x') & = & -\delta (x - x') \Gamma_{\alpha} (x) \int d^4 x''
D^{\alpha\beta} (x - x'') <\overline\psi (x'')\Gamma_{\beta} (x'') 
\psi (x'') >  \\
& & - \Gamma_{\alpha} (x) D^{\alpha\beta} (x - x') <\psi (x) \overline\psi
(x') > \Gamma_{\beta} (x')\,, \nonumber
\end{eqnarray}
and
\begin{eqnarray}
\Delta (x - x') & = & - \Gamma_{\alpha} (x) D^{\alpha\beta} (x - x')
< \psi (x) \overline\psi_T (x') > 
B\Gamma^T_{\beta} (x') B^{\dagger}\,,
\end{eqnarray}
where the equation for $\overline\Delta$ can be obtained using the Hermiticity
condition of Eq. (\ref{herm}).
These expressions become self-consistency equations
when we evaluate the expectation values by using their relationship to
the quasi-particle propagator,
\begin{eqnarray}
iS_F (x - x') = i 
\left( \begin{array}{cc}
\ G(x-x') & F(x-x') \\
 \widetilde F(x-x') & \widetilde G(x-x')
\end{array} \right)
= \left< \left( \begin{array}{c} \psi (x) \\ \psi_T(x) \end{array} \right)
\left( \overline\psi (x') \,, \overline\psi_T (x')\right) \right>\,,
\end{eqnarray}
which is itself a function of the mean fields. 

The equations in momentum space are
obtained by Fourier transforming the above expressions, giving
\begin{eqnarray}
\Sigma(k) &=& \Gamma_{\alpha}(0) D^{\alpha \beta}(0)
\int\frac{d^4 q}{(2 \pi)^4 }\mbox{Tr}\left[\Gamma_{\beta}(0) G(q)\right]
e^{i q^0 0^+} \\ \label{hfbks}
     & & \qquad\qquad -\int\frac{d^4 q}{(2 \pi)^4 } \Gamma_{ \alpha}(q)
D^{\alpha \beta}(q) G(k-q)\Gamma_{\beta}(-q)\,, \nonumber
\end{eqnarray}
and
\begin{eqnarray}
\Delta(k) &=& - \int\! \frac{d^4 q}{(2 \pi)^4 }\; 
 \Gamma_{\alpha}(q) D^{\alpha \beta}(q) F(k-q) 
B\Gamma^T_{\beta}(-q) B^{\dagger}  \,. \label{hfbkp}
\end{eqnarray}
To complete the set of equations, we include that constraining the
baryon density,
\begin{eqnarray}
\rho_B = <\overline\psi\gamma_0\psi > = \int\frac{d^4 q}{(2 \pi)^4 }
\mbox{Tr}\left[\gamma_0 G(q)\right] e^{i q^0 0^+}\,.
\end{eqnarray}
Solving the self-consistency equations in conjunction with this constraint,
we obtain the nonperturbative self-energy and pairing 
fields.

Here, we restrict our attention to $^1$S$_0$ pairing in symmetric 
nuclear matter.
The Hermiticity and transposition invariance conditions of Eqs. (\ref{herm}) 
and (\ref{tran}),
as well as the requirements of invariance under Lorentz and parity
transformations reduce the possible form of the self-energy field to
\begin{eqnarray}
\Sigma(k) & = & \Sigma_S(k) - \gamma_0\,\Sigma_0(k) + 
\vec\gamma\cdot\vec k\,\Sigma_V(k)\,, 
\end{eqnarray}
while we take the form of the pairing field to be 
\begin{eqnarray}
\Delta(k) = \bar\Delta(k) = (\Delta_S(k) - \gamma_0\,\Delta_0(k)
     - i\gamma_0\vec\gamma\cdot\vec k\,\Delta_T(k)) \vec\tau\cdot\hat n \,.
\end{eqnarray}
Both the self-energy and pairing terms have Lorentz scalar and
time-like vector components. To be consistent with our assumption of
symmetric nuclear matter, the self-energy must be an isoscalar.
The transposition invariance condition of Eq. (\ref{tran}) forces the scalar
pairing field to be an isovector. We have simplified the form of the pairing
field by assuming that it can be taken to be real and that the isospin
dependence can be isolated in an overall factor of $\vec\tau\cdot\hat n$,
where the unit vector $\hat n$ is arbitary.
The special case $\vec\tau\cdot\hat n = \tau_2$ corresponds to the 
standard one of proton-proton and neutron-neutron pairing. Although we
have not studied more elaborate isospin
dependences, we have examined complex solutions to the pairing equations
and found them to differ by only an overall phase from the solutions restrained
to be real. 

After substituting the simplified expressions for the mean fields
into the propagator and the self-consistency equations, the latter
can be reduced to coupled equations for the components of the mean
fields and the chemical potential, $\mu$. The self-consistency equations
contain contributions from both the negative-energy and the 
positive-energy states (from the Dirac sea and the Fermi sea).
To avoid the complications of renormalization, we have
discarded the poles of the HFB propagator corresponding to negative-energy
states. This procedure is not equivalent to the neglect of the HF 
negative-energy states performed in Ref. \cite{kring}, nor are the differences
small, as one might first expect. The contribution to the HFB states of
the negative-energy HF states reduces the magnitude of the pairing fields,
much as the contribution to the HF states 
of the negative-energy free states reduces the attractive scalar component
of the HF mean field. More details can be found in Ref. \cite{gcf},
where we compare the two approximations and show that discarding the HFB 
negative-energy states yields much more reasonable results.
  
\section*{Numerical results and discussion}

We have performed calculations of $^1$S$_0$ pairing in symmetric nuclear
matter for various sets of interaction parameters (meson-nucleon coupling 
constants and meson masses\cite{sw,hs-hf,bmgm}).
We introduced, as an additional
parameter, a momentum cutoff at $|\vec k| = \Lambda$ (in the nuclear matter 
rest frame), which limits the momentum integrations in the self-consistency
equations. Such a cutoff could be considered a crude approximation to the 
nucleon-meson vertex form factors that our calculations do {\em not}
contain. We performed calculations for various values of the momentum cutoff.

As the self-consistency equations for the pairing fields are proportional
to the pairing fields, it is always possible to find a solution in which
these fields are null. This zero pairing-field solution is just the normal 
HF one. In our calculations, we have found this to be the only 
self-consistent solution at sufficiently high baryon density. At densities
lower than about two thirds of the saturation density,
$\rho_B\leq 2\rho_{B0}/3$, we also find a non-trivial HFB solution. (We will
not consider the region of exponentially small pairing fields discussed in
Ref. \cite{khodel}.)

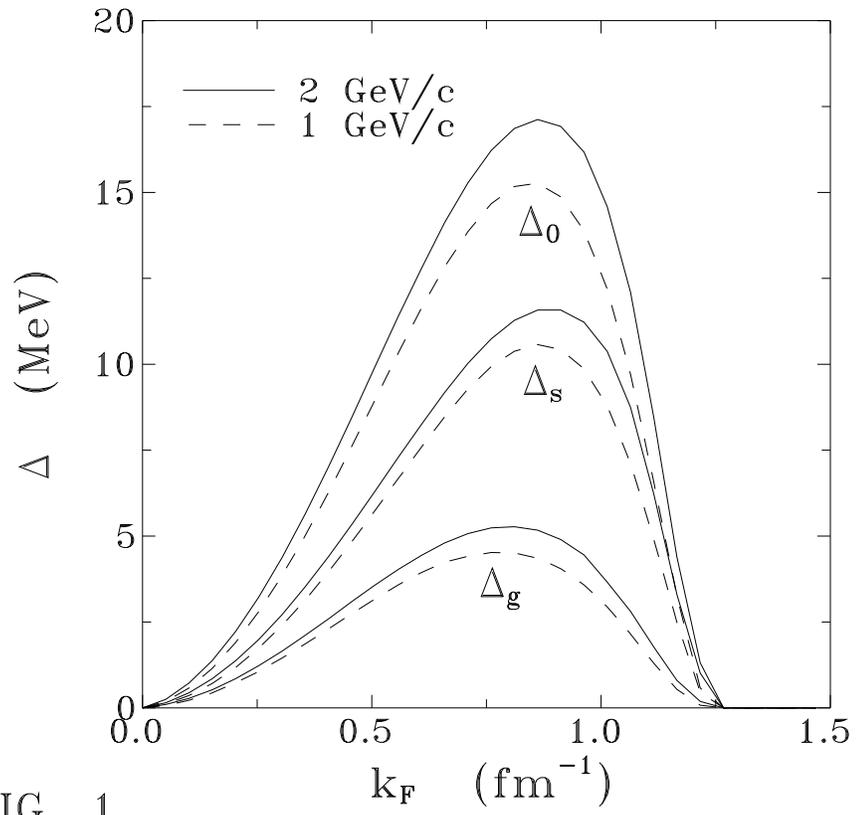
\begin{figure}
\caption{ Pairing fields and gap functions at the Fermi momentum,
obtained with the parameters of Ref.~[19] at two values of the
cutoff parameter, $\Lambda=2$ GeV/c (solid lines) and $\Lambda=1$ GeV/c
(dashed lines), for the $\sigma$-$\omega$ model. The fields are
$\Delta_0$, $\Delta_s$, and $\Delta_g$, are in decreasing order.}
\label{fig1}
\end{figure}

We display, in Fig.\ \ref{fig1}, the two principal components of the pairing
field, $\Delta_S(k_F)$ and $\Delta_0(k_F)$, evaluated at the Fermi momentum,
obtained for two different values of the momentum cutoff $\Lambda$ using the
$\sigma$-$\omega$ model parameters of Ref. \cite{bmgm}. The third 
component of the pairing field, $\Delta_T$, is three orders of magnitude 
smaller than $\Delta_S$ and $\Delta_0$ and is not shown. The components of
the pairing field are plotted as functions of the Fermi momentum, where the 
latter is {\em defined} through its standard relation to the baryon density, 
that is, $\rho_B=\gamma k_F^3/6\pi^2$  where $\gamma$ is 2 for nuclear
matter and 1 for neutron matter. We observe that the
magnitudes of the components of the pairing field, represented here by their
values at the Fermi momentum, $\Delta_{0,s}(k_F)$, depend strongly on
the Fermi momentum (and thus the baryon density). In both of the cases shown,
which are typical ones, the fields increase rapidly at small densities,
reach a maximum at about $1/4$ the saturation density, and fall back to
zero before the saturation density is reached. We observe that
the magnitude of the components of the pairing field also clearly depend
on the value of the momentum cutoff. The form of the functional
dependence of the fields on the Fermi momentum, however, is fairly insensitive
to the value of the cutoff.

Diagonalizing the equation of motion, Eq. (\ref{eqmot}), in the 
particle-particle and hole-hole subspaces --- equivalent to expanding
in the HF basis in these subspaces --- reduces the pairing fields to  effective
ones for the positive- and negative-energy states and to a coupling term which
is small for low momenta (but increases with the momentum)\cite{gcf}. The
effective pairing field for the positive-energy states, which we call the gap
function, is   
\begin{eqnarray}
\Delta_g(k) = \frac{M^{\star}(k)}{E^{\star}_k} \Delta_0(k) - \Delta_S(k)
              +i\frac{k^{\star}\mid\vec k\mid}{E^{\star}_k}\Delta_T(k)\,,
\end{eqnarray}
where $k^{\star}=(1+\Sigma_V(k))|\vec k|$, $\,M^{\star}=M+\Sigma_0(k)$, and
$E^{\star}=\sqrt{k^{\star 2} + M^{\star 2}}$.
This is the quantity whose role is closest to that of the non-relativistic
pairing field. Like the non-relativistic self-energy, it too
is the difference between two larger relativistic quantities. This is evident
in Fig.\ \ref{fig1}, in which the gap functions for the two different values
of the momentum cutoff are displayed together with the components of the
pairing field. We observe that different values of the momentum cut-off result
in gap functions $\Delta_g$ of different magnitude but with approximately the
same form of functional dependence on the Fermi momentum. The gap
functions are quite similar, in form, to those obtained in various
non-relativistic calculations using realistic potentials 
\cite{baldo1,khodel,nordic,vgdick}.

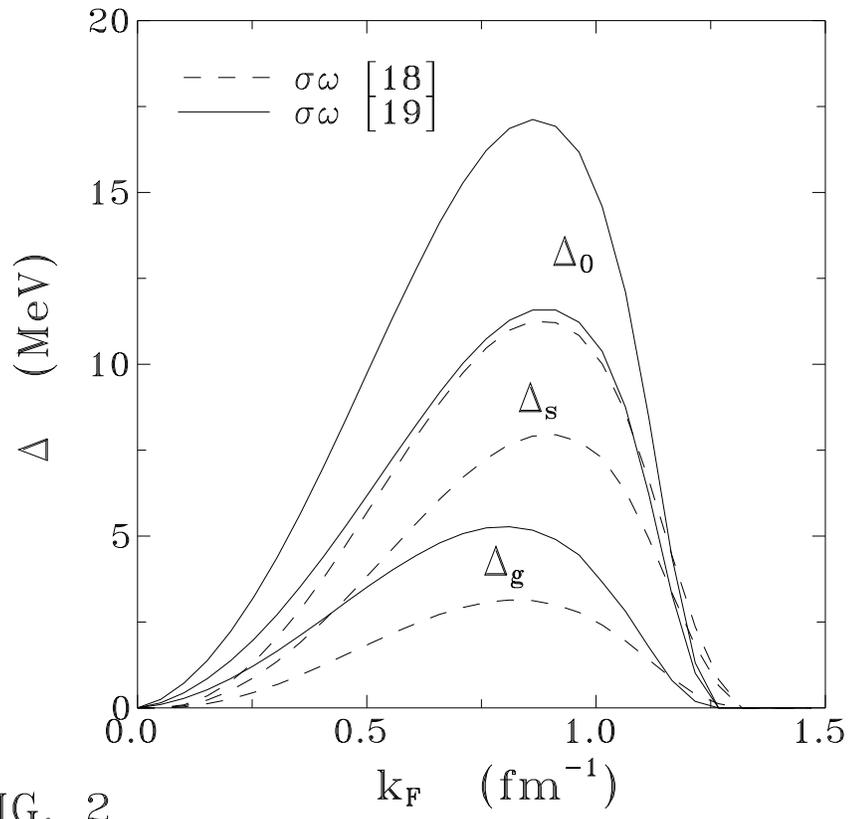
\begin{figure}
\caption{ Pairing fields and gap functions at the Fermi momentum,
obtained with a cutoff parameter of $\Lambda=2$ GeV/c using the parameters
of Ref.~[18] (dashed lines) and Ref.~[19] (solid lines),
for the $\sigma$-$\omega$ model. The fields
$\Delta_0$, $\Delta_s$, and $\Delta_g$, are in decreasing order.}
\label{fig2}
\end{figure}

In Fig.\ \ref{fig2}, we compare the components of the pairing field and
gap functions obtained using the parameters of Ref. \cite{hs-hf} and
Ref. \cite{bmgm}, at a fixed value of the cutoff momentum $\Lambda$,
again in a $\sigma$-$\omega$ model. Here, we 
observe that different values of the interaction parameters also result in
pairing fields $\Delta_0$ and $\Delta_s$ and gap functions $\Delta_g$ of
different magnitudes but, again, with roughly the same form of functional
dependence on the Fermi momentum.

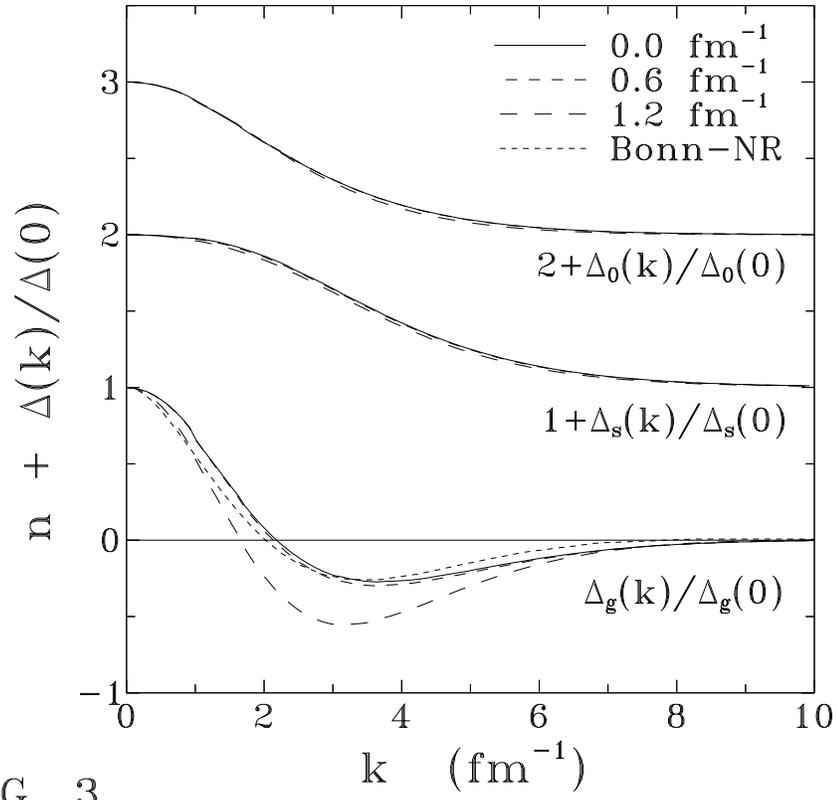
\begin{figure}
\caption{ Normalized pairing fields and gap functions, as a 
function of the momentum, for several values of the Fermi momentum,
obtained with the $\sigma$-$\omega$ model parameters of Ref.~[19]
and a cutoff parameter of $\Lambda=2$ GeV/c. The ratio 
$ t_{10}(k,0;0)/t_{10}(0,0;0)$ of the Bonn-B $^1$S$_0$ half-on-shell T-matrix
at zero energy is also shown (short-dashed line).}
\label{fig3}
\end{figure}

As we expect the pairing fields and the gap function, like the
two-nucleon vertex function, to contain only short-range
two-nucleon correlations, we expect their dependence on the momentum
to be approximately the same over a wide range of nuclear densities. We
would expect significant changes in their momentum dependence only at
densities high enough for the probability to become appreciable of a third
nucleon to be found simultaneously with the pair within the effective
radius of about 0.5 fm that is characteristic of the short-range correlations.
In Fig.\ \ref{fig3}, we
show the normalized momentum dependence of the pairing fields and gap 
function (at a fixed value of the Fermi momentum), $\Delta(k)/\Delta(0)$, 
obtained using the interaction parameters of Ref. \cite{bmgm}
and a momentum cutoff of $\Lambda= 2$ GeV/c, for several values of
the baryon density. The curve corresponding to zero density gives the
momentum dependence of the virtual state solution to the vacuum pairing
equations, which is discussed below. Confirming our expectations, at
low density, we find that the momentum dependence of the pairing fields
and gap function is indeed almost independent of the density. As the density
increases, differences begin to appear in the gap function but remain small
almost up to the density at which pairing disappears. The momentum dependence
of the pairing fields remains almost identical to that of the virtual state
over the entire range of densities for which pairing occurs.

We note that in Ref. \cite{khodel} a similar comparison was made of the
momentum dependence of the non-relativistic gap function at various values
of the baryon density, but for the case of neutron matter. They observed
slightly larger variations of the normalized gap function than the ones we have
found but, as their calculations were non-relativistic, did not observe the 
invariance of the pairing fields. Our results for neutron matter are almost
identical to those in Fig.\ \ref{fig3} and are not shown here. We want to 
emphasize, however, that the momentum dependence of the neutron matter pairing
fields is just as independent of the matter density as is the momentum
dependence of the pairing fields in the case of nuclear matter.

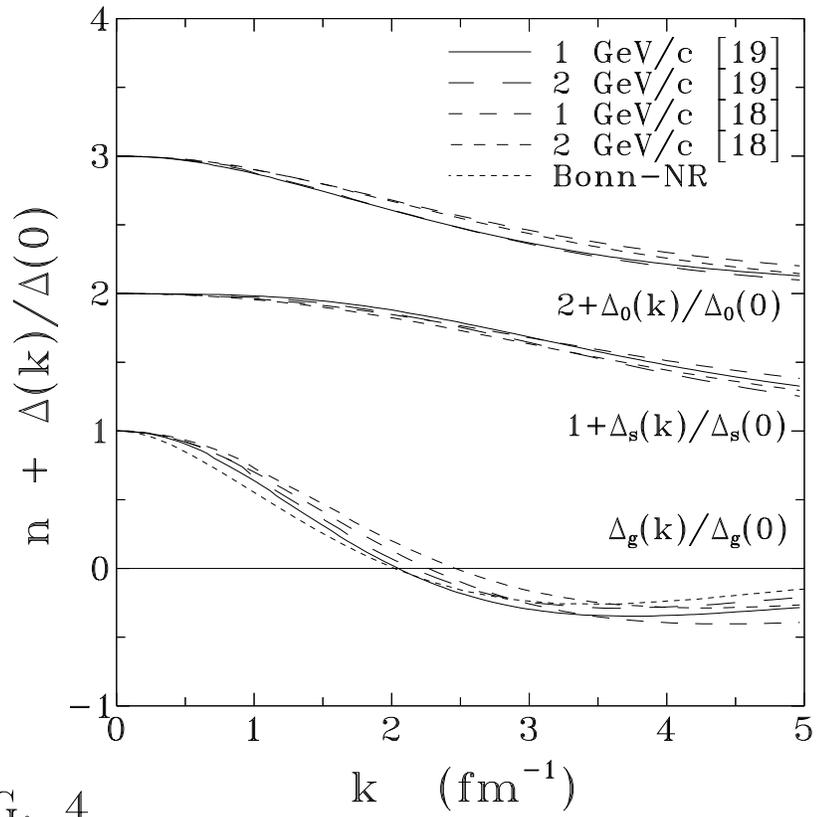
\begin{figure}
\caption{ Normalized pairing fields and gap functions, as a 
function of the momentum, at a Fermi momentum of $k_F=0.5$ fm$^{-1}$,
obtained with the parameters of Ref.~[18] at
two values of the cutoff parameter, $\Lambda=2$ GeV/c (solid lines) and
$\Lambda=1$ GeV/c (dashed lines), and the parameters of Ref.~[19]
with a cutoff of $\Lambda=1$ GeV/c (dotted line),
for the $\sigma$-$\omega$ model. The ratio 
$ t_{10}(k,0;0)/t_{10}(0,0;0)$ of the Bonn-B $^1$S$_0$ half-on-shell T-matrix
at zero energy is also shown (short-dashed line).}
\label{fig4}
\end{figure}

We would also like the short-range correlations contained in the pairing fields
and gap function to be relatively independent of the interaction parameters,
at least if we believe that these parameters provide a reasonable description
of the actual physical situation. We can see in Fig.\ \ref{fig4} that, to a
certain extent, this
is the case. There we show the normalized momentum dependence of the pairing
fields and gap function, obtained for two different values of the momentum
cutoff using the $\sigma$-$\omega$ model parameters of Ref. \cite{hs-hf} and of
Ref. \cite{bmgm}.
We see that the normalized momentum-dependence of the pairing fields
and gap function, although similar, is clearly different for the two sets
of interaction parameters. However, the momentum-dependence of
the fields depends only weakly on the values of the cutoff, $\Lambda$, as
is also the case for the other sets of interaction parameters that we
have used.

In the limit of zero density, the integral equation for the pairing field,
Eq. (\ref{hfbkp}), reduces to the ladder approximation to the
Bethe-Salpeter equation for a two-particle bound state, in which the
the chemical potential is identified with the two-body bound-state energy
and antiparticle propagation in the two-nucleon center-of-mass frame has been 
discarded. For a physically reasonable interaction, this has no 
nontrivial solution in the $^1$S$_0$ channel, as the two-nucleon system has
no bound state there. The $^1$S$_0$ virtual state appears as a solution when 
the equation is analytically extended to the second energy sheet. As we have
already stated, the energy of the virtual $^1$S$_0$-state is extremely small,
so that its vertex function dominates the short-range correlations
in two-nucleon scattering at low energies. As the virtual state and the pair
state at low nuclear matter densities are solutions of essentially the same
equation (and contain essentially the same physics), we expect the short-range
correlations of the pair state to be similarly dominated by the short-range
correlations of the virtual state. We thus expect the the components of
the pairing field to have a momentum dependence similar to that of the
components of the vertex function of the virtual state. We have seen in 
Fig.\ \ref{fig3} that, at low densities, this is indeed the case. As
the gap function $\Delta_g$ has a momentum dependence similar to that of
the non-relativistic vertex function of the virtual state $d_v(q)$ we expect,
as discussed above, that it should also have a dependence on the momentum
similar to that of the half-on-shell matrix element of the zero-energy
non-relativistic $^1$S$_0$ T-matrix, $t_{10}(q,0;0)$. That is, we expect that
\begin{eqnarray}
\frac{\Delta_g(q)}{\Delta_g(0)}\approx\frac{d_v(q)}{d_v(0)}
\approx \frac{t_{10}(q,0;0)}{t_{10}(0,0;0)} \;.
\end{eqnarray}
In Figs.\ \ref{fig3} and \ \ref{fig4}, we have plotted, together with
the normalized pairing fields, the normalized non-relativistic
zero-energy half-on-shell $^1$S$_0$ T-matrix, for the
case of the Bonn-B potential\cite{mach3}, which includes the exchange of
$\sigma$, $\omega$, $\rho$, $\pi$, $\eta$ and $\delta$ mesons
between nucleons. We see in Fig.\ \ref{fig3}, for the interaction parameters
of Ref. \cite{bmgm}, that the agreement between the ratios of gap functions 
and half-on-shell T matrices is quite good. We observe in Fig.\ \ref{fig4} 
that the ratios of gap functions for the interaction parameters of Ref. 
\cite{hs-hf} do not agree quite as well as those of Ref. \cite{bmgm} with the
ratios of half-on-shell T matrices, although they are still quite similar, as
are the gap functions of the other sets of interaction parameters.
We thus conclude that the momentum dependence of the effective interactions
that we used does provide a reasonable description of the actual physical
situation.

In contrast with the almost unchanging momentum dependence of the pairing
fields, we find that their magnitudes depends strongly on the interaction
parameters and the value of the cutoff. This reflects the fact that
nuclear pairing is the result of a delicate balance between short-range
repulsion and long-range attraction.\cite{baldo3p} Changing the interaction
parameters or the momentum cutoff changes the point of this balance,
modifying the magnitudes of the fields. However, such modifications 
do not appear in the magnitudes of the pairing fields alone.
As we have noted, in the zero
density limit, the equation for the pairing field, Eq. (\ref{hfbkp}),
reduces to the ladder approximation to the Bethe-Salpeter equation for a 
two-particle bound state in the $^1$S$_0$ channel. For some sets
of interaction parameters and momentum cutoff, a bound state can indeed be 
found. In other cases, where no bound state exists, analytic continuation 
of the pairing equations to the second energy sheet reveals a virtual state
in the $^1$S$_0$ channel.

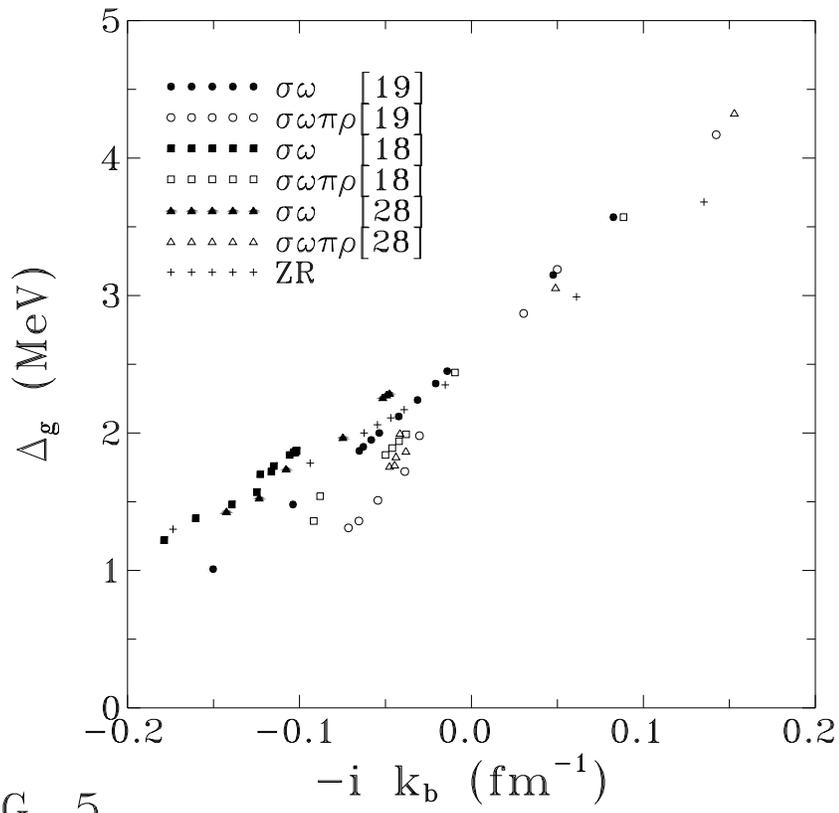
\begin{figure}
\caption{ The gap function $\Delta_g$ at a Fermi momentum of 
$k_F=0.5$ fm$^{-1}$ plotted versus the momentum of the bound or virtual 
state, $-i\, k_b$, for each set of interaction parameters and cutoff.
Calculations with $\sigma$ and $\omega$ mesons and $\sigma$, $\omega$,
$\pi$ and $\rho$ mesons are labelled in the
figure. The zero-range calculation is labelled ZR.}
\label{fig5}
\end{figure}

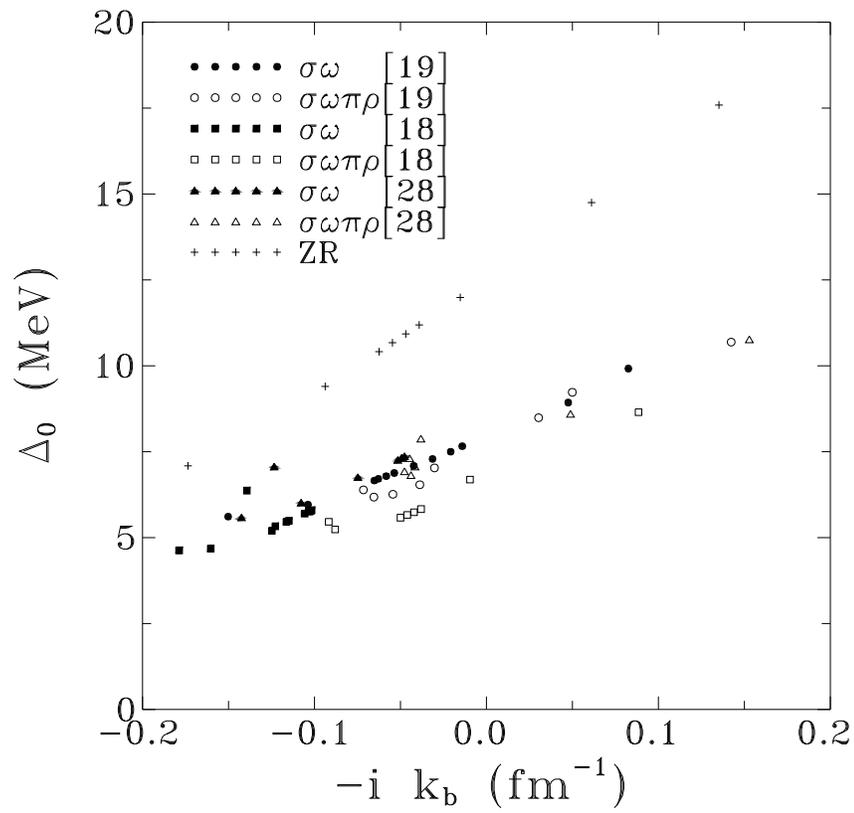
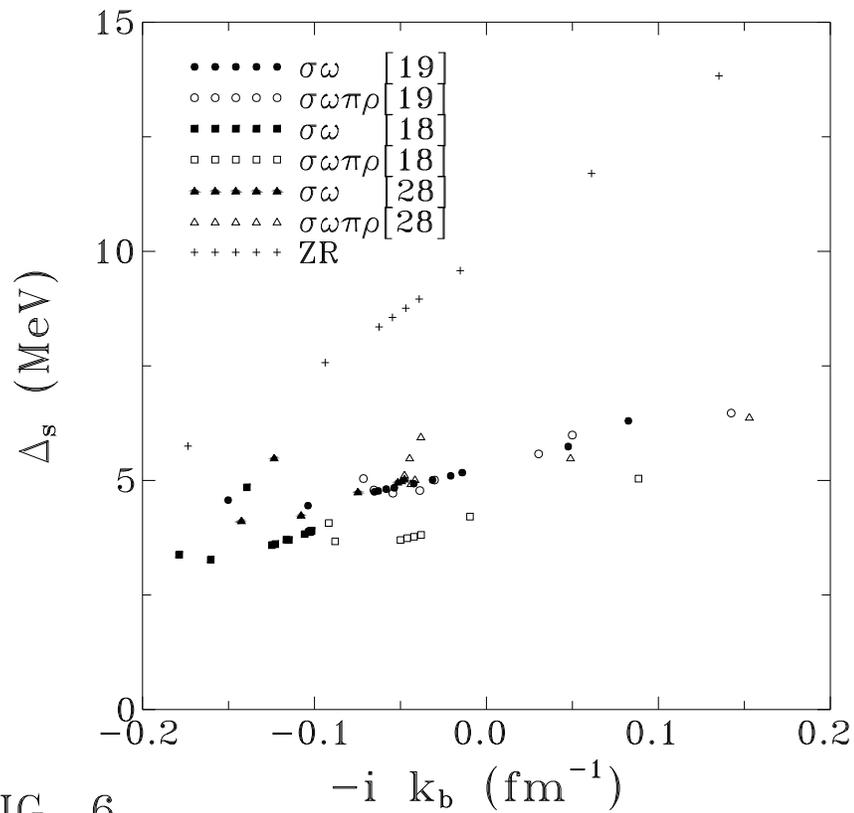
\begin{figure}
\caption{ The pairing fields $\Delta_0$ and $\Delta_S$ at a Fermi
momentum of 
$k_F=0.5$ fm$^{-1}$ plotted versus the momentum of the bound or virtual 
state, $-i\, k_b$, for each set of interaction parameters and cutoff.
Calculations with $\sigma$ and $\omega$ mesons and $\sigma$, $\omega$,
$\pi$ and $\rho$ mesons are labelled in the
figure. The zero-range calculation is labelled ZR.}
\label{fig6}
\end{figure}

We find the magnitudes of the gap function  and the pairing fields
at low densities to be strongly correlated with the 
location in the complex momentum plane of the virtual state
or bound state in the vacuum that is produced by each set of interaction 
parameters + cutoff. We display these correlations in Figs.\ \ref{fig5}
and \ \ref{fig6}, where comparison is made between several different
parameter sets with varying numbers of mesons. Calculations were
performed including the $\sigma$, $\omega$, $\rho$ and $\pi$ mesons and
also including just the $\sigma$-$\omega$ mesons, as the latter dominate the
gross features of the pairing fields and gap function.
Calculations were also performed using a relativistic zero-range interaction,
in which the $\sigma$ and $\omega$ mesons are taken to be infinitely heavy and
the coupling constants are defined so as to yield the correct nuclear-matter
saturation point. For each parameter set, the momentum cutoff $\Lambda$ was
varied with all other parameters kept fixed. For each value of the momentum
cutoff, we determined the pairing gap at $k_F$=0.5 fm$^{-1}$, about 1/8 of
the saturation density, and obtained the bound or virtual state energy in
the vacuum.

We observe in Figs.\ \ref{fig5} and \ \ref{fig6} that the value of the
free two-nucleon bound or
virtual state energy essentially determines the magnitudes of the gap function
$\Delta_g(k_F)$ and the pairing fields $\Delta_{0,s}(k_F)$ at
$k_F$=0.5 fm$^{-1}$. The $\sigma$-$\omega$ results agree well among
themselves, with their values for each of the pairing fields and gap 
function all lying within a narrow band. Although the inclusion of other
mesons increases the size of the bands, their widths still remain small
relative to the magnitudes of the fields. In Fig.\ \ref{fig5}, we see that
even the zero-range
model yields results for the gap function in good agreement with those of the
other calculations. However, it cannot reproduce the values of the components
of the pairing field, $\Delta_{0,s}(k_F)$, obtained with the finite-range
models, as is apparent in Fig.\ \ref{fig6}.

The correlations between the components of the pairing field,
$\Delta_{0,s}(k_F)$, and the position of the virtual state, shown in
Fig.\ \ref{fig6}, yield two distinct lines. In each case, one of
these lines contains most of the finite-range results while the other contains
the zero-range calculations. Several of the finite-range points lie
between the two lines. The finite-range calculations that fall 
close to the zero-range line all correspond to extremely low values of the
momentum cutoff, $\Lambda\lesssim 500$ MeV/c. They are thus similar to the
momentum-independent zero-range results in the sense that the pairing fields
(but not the gap function) in these cases vary little as a function of the
momentum. (Better said, they have very little room in which to vary.) This
is clear from the leftmost portions of Figs. \ \ref{fig3} and \ \ref{fig4},
in which the momentum dependence expected of the pairing fields 
for values of the cutoff $\Lambda\lesssim 500$ MeV/c is displayed in the range
of values $k\lesssim 2.5$  fm$^{-1}$.

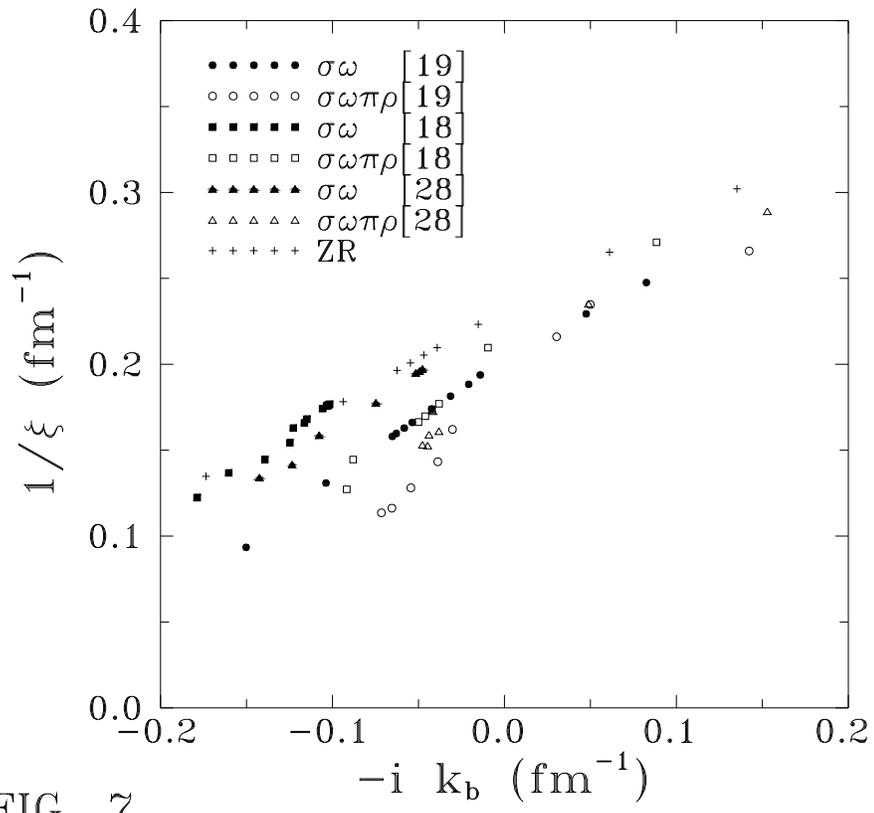
\begin{figure}
\caption{ The inverse of the coherence length $1/\xi$ at a Fermi
momentum of 
$k_F=0.5$ fm$^{-1}$ plotted versus the momentum of the bound or virtual 
state, $-i\, k_b$, for each set of interaction parameters and cutoff.
Calculations with $\sigma$ and $\omega$ mesons and $\sigma$, $\omega$,
$\pi$ and $\rho$ mesons are labelled in the
figure. The zero-range calculation is labelled ZR.}
\label{fig7}
\end{figure}

We can understand better the correlation between the magnitudes of the
pairing fields and the location of the virtual state if we examine the
coherence length of the pairing fields. The coherence length $\xi$ of a
bound pair is defined as its root mean square radius. In terms of the
momentum-space wave function of the pair, $\chi(\vec k)$, we have
\begin{eqnarray}
\xi^2=\int\,d^3 k\, \frac{\partial\chi^{\dagger}(\vec k)}{\partial\vec k}
\frac{\partial\chi(\vec k)}{\partial\vec k}\left/\int\, d^3 k\,
\chi^{\dagger}(\vec k) \chi(\vec k)\right.\,.
\end{eqnarray}
In the Bethe-Salpeter equation, the pair wave function can be expressed in
terms of
the residue of the two-particle propagator at its pole. By analogy, we argue
that here we can obtain the (un-normalized) pair wave function as the residue
of the anomalous propagator $F(k)$ at its pole,
\begin{eqnarray}
\chi(\vec k) = \frac{1}{2\pi i}\int\,d\omega F(\vec k,\omega) e^{i\omega 0^+}
\,,
\end{eqnarray}
where we retain only the contribution of the positive-energy pole, to be
consistent with the vacuum truncation discussed earlier. In Fig.\ \ref{fig7},
we plot the inverse of the coherence length, $1/\xi$, at $k_F$=0.5 fm$^{-1}$,
as a function of the position of virtual or bound state in the vacuum, just as 
was done in Figs.\ \ref{fig5} and \ \ref{fig6}. Here too, we observe a similar
strong correlation. What is most important at the moment, however, are the
typical values obtained for the coherence length. We observe that, in the range
of calculations shown, these never fall below 3 fm. Near the physical position
of the virtual state, the value of the coherence length lies between about
4 to 10 fm. As will be seen shortly, the Fermi momentum at which we have 
shown the various correlations is close to that for which the coherence
length of the pair attains its minimum value. Thus we can claim that the 
coherence length of the physical bound pair never becomes smaller than
about 4 fm.

We can thus justify the correlations observed with the following argument:
For the parameter sets of physical interest,
the size of a bound pair at low density is much greater than the range of
the interaction. As the short-range pair correlations are about the same
for all sets of finite-range interaction parameters, it is then a single
parameter, the net attraction, that determines the differences in the binding
energy of the pair obtained with each of the parameter sets.
It suffices to fix the size (or energy) of the pair at one value of the
density to obtain its trend over the entire range of values for which the 
details of the interaction can be neglected. We choose to fix the net 
attraction between pairs by fixing the position in the complex momentum plane 
of the $^1$S$_0$ virtual state in the vacuum. As we have seen, this is 
equivalent to fixing the two-nucleon singlet scattering length.

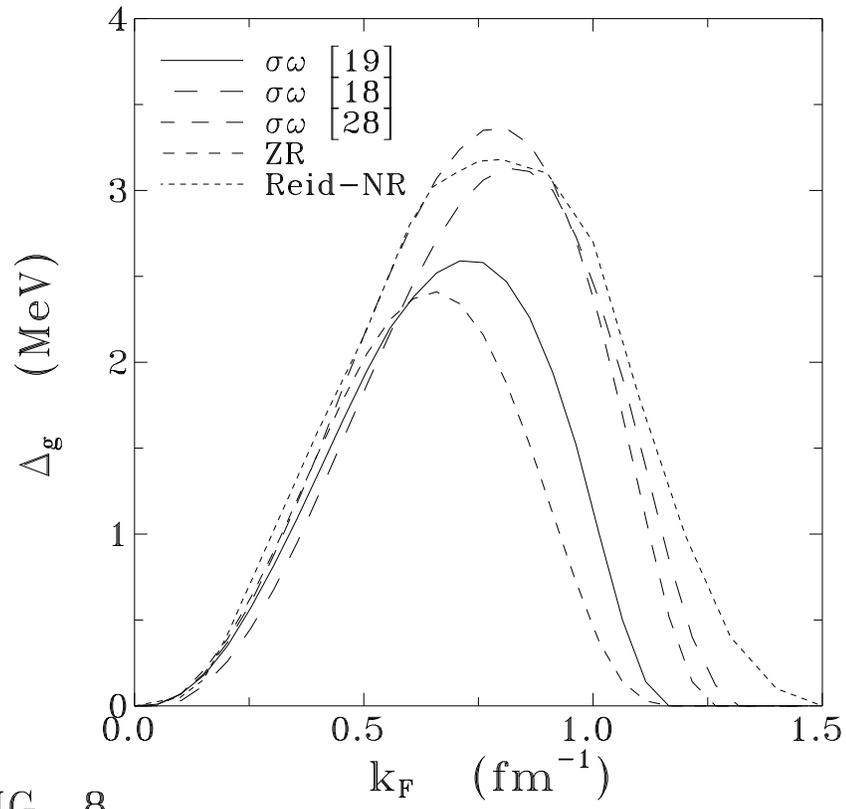
\begin{figure}
\caption{ Gap functions $\Delta_g$ obtained for several sets of 
$\sigma$-$\omega$ interaction
parameters that have been constrained so as to yield a virtual state close 
to the physical one. The non-relativistic calculation of Ref.~[30],
which uses a Reid soft-core potential, is also shown.}
\label{fig8}
\end{figure}

We thus fix a momentum cutoff for
each parameter set so that it places the $^1$S$_0$  virtual state at its
physical location of $k_v \approx -0.05 i $ fm$^{-1}$. This is impossible
for the $\sigma$-$\omega$ interaction parameters of Ref. \cite{hs-hf}, for
which $i k_v \gtrsim 0.08 $ fm$^{-1}$ for $\Lambda\gtrsim 500$ MeV/c. In this
case, we fix the virtual state at $k_v \approx -0.08 i $ fm$^{-1}$, its
closest point to the physical location. The pairing fields that result are
then fairly consistent among themselves and with the non-relativistic result
of Ref. \cite{vgdick} for low values of the Fermi momentum, as can be seen in
Fig.\ \ref{fig8}. Our results are also very similar to those obtained
in many other non-relativistic calculations of both nuclear matter and
neutron matter (for which the gap function is from 10\% to 20\%larger), which
are not shown here \cite{baldo1,khodel,nordic}. Although the calculations
shown are restricted to those using the $\sigma$-$\omega$ parameter sets, 
the gap functions obtained using the $\sigma$-$\omega$-$\pi$-$\rho$ interaction
parameters are quite similar. Even the zero-range model can describe the 
behavior of the Fermi-momentum gap-function at low densities, when it used
with an appropriate cutoff $\Lambda$.

We note that the good agreement between different sets of interaction
parameters does not extend up to densities at which the magnitude of
the pairing fields begins to decrease. There, differences in the details of
the interplay between attraction and short-range repulsion become important, 
introducing a more elaborate dependence on the parameter sets. To the extent
to which these details are reflected in the short-range correlations, that is
to say, in the momentum dependence of the pairing fields, we can compare this 
momentum dependence with that of the half-on shell singlet T-matrix to choose
the most ``physical'' parameter sets. As we have not performed relativistic
calculations of the T-matrix, we have compared the non-relativistic
calculations  using the Bonn potential
to our gap functions to conclude that the $\sigma-\omega$ and
$\sigma-\omega-\pi-\rho$ parameter sets of Ref.\ \cite{bmgm} are equally good
and are the most ``physical'' of those we have studied. The half-on-shell
singlet T-matrix obtained using the Bonn potential and the gap function
obtained using the $\sigma-\omega$ parameters of Ref. \cite{bmgm} are shown
in Figs.\ \ref{fig3} and \ref{fig4}.

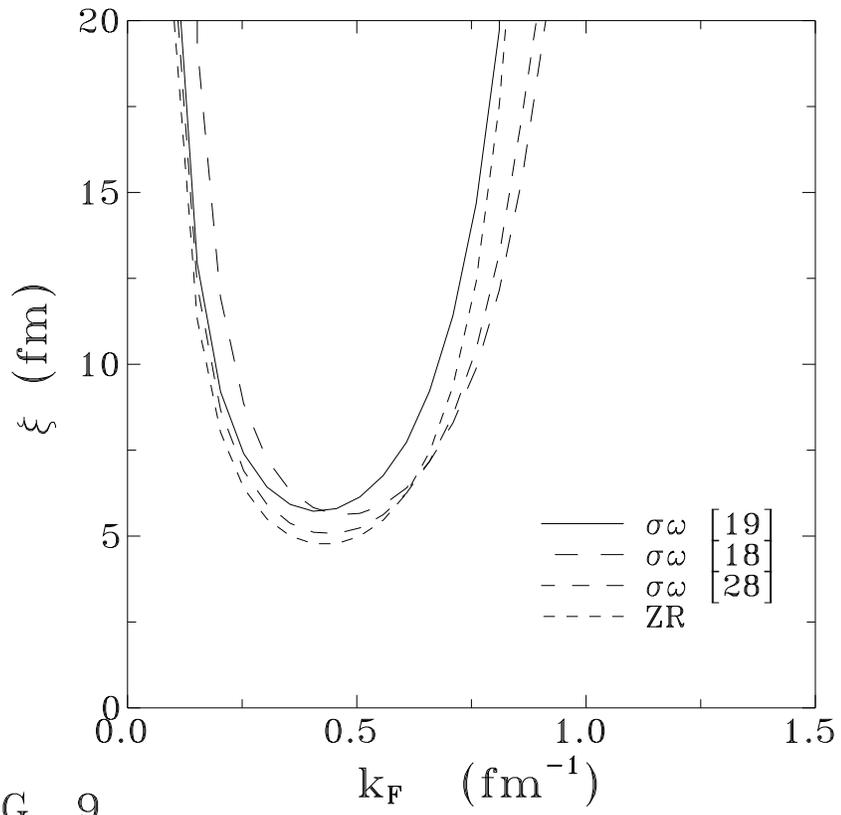
\begin{figure}
\caption{ Coherence lengths $\xi$ obtained for the same sets of 
$\sigma$-$\omega$ interaction parameters constrained so as to yield a virtual
state close to the physical one as in Fig. 8.}
\label{fig9}
\end{figure}

We show in Fig.\ \ref{fig9} the coherence lengths $\xi$ obtained using the
physical values of the momentum cutoff that resulted in the gap functions of
Fig.\ \ref{fig8}. We observe that the coherence length drops rapidly from
infinity, at low densities, to reach a minimum of about 5 fm at about 1/10 the 
density of saturated nuclear matter. It rises again rapidly, becoming infinite 
at the density at which pairing disappears. The minimum coherence length is 
fairly independent of the interaction parameters and is about 5 or 6 fm in all
cases studied. The coherence length in neutron
matter displays a very similar dependence on the Fermi momentum $k_F$ but
is about 10\% smaller than in nuclear matter of the same Fermi momentum.

We conclude by interpreting the spatial form of a bound $^1$S$_0$ pair in 
nuclear matter, when it exists, in terms of the magnitude and form of the 
pairing field, $\Delta(k)$.
The short-range correlations of the pair, characterized by the
form of the pairing field, $\Delta(k)/\Delta(0)$, are almost independent of 
model parameters, momentum cutoff and nuclear density, and are almost identical
to those of the two-nucleon $^1$S$_0$ virtual state in the vacuum. The
asymptotic nature of the pair, is roughly determined by the magnitude of
the pairing fields at the Fermi momentum, which are strongly varying 
functions of the nuclear matter density. Based on the variations observed
in the pair coherence length, we estimate
that a $^1$S$_0$ pair in nuclear matter is most tightly bound at about $1/10$
the nuclear saturation density, somewhat below the density for which the
pairing fields reach their maximum values. For larger and smaller values of
the density, the pair is larger in extent and less bound.

\section*{Final remarks}

Our numerical results indicate that the two principal physical
ingredients that fix the form and magnitude of the pairing fields and
the gap function are the energy of the nucleon-nucleon virtual state and
the momentum dependence of its vertex function or, equivalently, the
two-nucleon singlet scattering length and the momentum dependence of 
the half-on-shell zero-energy singlet T-matrix. The short-range two-nucleon
correlations contained in the momentum dependence of the pairing fields and
the virtual state are determined by the interplay between the short-range
repulsion and medium-range attraction of the nucleon-nucleon interaction.
If the short-range correlations are held fixed, the magnitude of the pairing
fields and the position of the virtual state depend only on the
overall strength of the interaction.

Pairing is also expected to occur in the $^3$S$_1$-$^3$D$_1$ deuteron
channel. This possibility has been investigated by two groups
\cite{baldo3,baldo3p,vgdick}, which both obtained nonrelativistic S-D gap
functions that reach values of about 10 MeV. Although we have not performed
such calculations, we note that the large values obtained for the S-D
gap function are not surprising, given the correlation we observed in
Fig.\ \ref{fig5}. Since the deuteron is a bound pair while the $^1$S$_0$ 
two-nucleon state is virtual, the stronger effective two-nucleon potential 
in the $^3$S$_1$-$^3$D$_1$ channel would also be expected to produce a much
larger gap function and pairing fields.

Several recent works have pointed out that medium polarization
effects result in an important renormalization of the effective
(non-relativistic) nucleon-nucleon interaction \cite{wamba1,wamba2,baldo5}.
One of the observed effects of this renormalization is a large reduction in
the magnitude of the gap function. Such a reduction is 
consistent with our results if the polarization effects of the medium
reduce the attraction of the effective two-nucleon potential. Another recent
study has shown that the quark substructure of the nucleons and exchanged
mesons is also expected to lead to a less attractive two-nucleon
potential\cite{mach4}. We plan to extend our study of pairing to next take into
account such effects.

\section*{Acknowledgements}

B.V. Carlson and T. Frederico acknowledge the support of the Conselho
Nacional de Desenvolvimento Cient\'{\i}fico e Tecnol\'ogico - CNPq, Brazil.

\end{document}